\documentclass[prb,twocolumn,superscriptaddress,showpacs]{revtex4}
\usepackage{amsmath}
\usepackage{bm}
\usepackage{graphicx}
\usepackage{amssymb}
\usepackage{color}
\begin{document}
\newcommand{\figwidth}{0.95\columnwidth}
\newcommand{\ffigwidth}{0.4\columnwidth}
\newcommand{\warwick}{Department of Physics and Centre for Scientific Computing, University of Warwick, Coventry,
  CV4 7AL, United Kingdom}
\newcommand{\usal}{Departamento de Fisica Fundamental, Universidad de Salamanca, 37008 Salamanca, Spain}
\title{Multifractal analysis of the metal-insulator transition in the 3D Anderson model I: Symmetry relation under typical averaging}
\author{Louella J. Vasquez}
\affiliation{\warwick}
\author{Alberto Rodriguez}
\affiliation{\warwick}
\affiliation{\usal}
\author{Rudolf A. R\"omer}
\email[Corresponding author: ]{R.Roemer@warwick.ac.uk}
\affiliation{\warwick}
%
%
\begin{abstract}
The multifractality of the critical eigenstate at the metal to insulator transition (MIT) in the three-dimensional Anderson model of localization is characterized by its associated singularity spectrum $f(\alpha)$.
Recent works in 1D and 2D critical systems have suggested an exact symmetry relation in $f(\alpha)$.
Here we show the validity of the symmetry at the Anderson MIT with high numerical accuracy and for very large system sizes. We discuss the necessary statistical analysis that supports this conclusion.
We have obtained the $f(\alpha)$ from the box- and system-size scaling of the typical average of the generalized inverse participation ratios.  We show that the best symmetry in $f(\alpha)$ for typical averaging is achieved by system-size scaling, following a strategy that emphasizes using larger system sizes even if this necessitates fewer disorder realizations.
\end{abstract}
\maketitle
\section{Introduction}
\label{sec-intro} 
The Anderson model of localization has been a subject of intense analytical \cite{And58,LeeR85,AbrALR79} and numerical \cite{KraM93,RomS03,EilFR08} studies  for decades.  Anderson in his seminal paper\cite{And58} had demonstrated that, at absolute zero temperature and in the absence of external fields and electron-electron interactions, a sufficiently strong disorder can drive a transition from a metallic to an insulating state (MIT).
The scaling theory of localization\cite{AbrALR79} has shown that such a transition arises generically for systems with dimension $d>2\,$.

A characteristic feature of this critical transition is the strong multifractality of its wavefunction amplitudes.\cite{Aok86,SchG91,FalE95a,MorKMW03,MorKMG02}
The critical eigenstate being neither extended nor localized reveals large fluctuations of wavefunction amplitudes at all length scales.
The characterization of multifractality is most often given in terms of the singularity spectrum $f(\alpha)$.  It can be computed from the $q$-th moments of the inverse participation ratio $P_q$ which defines the scaling behaviour of $P_q\propto \lambda^{d(q-1)+\Delta_q}$ with length $\lambda$.\cite{Jan94a}
The anomalous exponents $\Delta_q$ determine the scale dependence of the wave function correlations \cite{EveMM08} and separate the critical point from the metallic phase (for which $\Delta_q=0$).
By carrying out such a multifractal analysis\cite{Jan94a, ChaJ89, MilRS97} (MFA) various critical properties can be obtained such as the critical disorder $W_c$,\cite{MilE07} the position of the mobility edges and the disorder-energy phase diagram,\cite{GruS95} as well as the critical exponents of the localization length.\cite{Kra93,PooJ91,YakO98}

From an analytical viewpoint not much is known about the singularity spectrum.  An approximate expression can be obtained in the regime of weak multifractality, i.e.\ when the critical point is close to a metallic behaviour. This applies to the Anderson
transition in $d=2+\epsilon$ dimensions with $\epsilon \ll 1$. In this case a parabolic dependence of the spectrum is found as
 	$f(\alpha) \simeq d-{\left[\alpha -(d+\epsilon)\right]^2}/{4\epsilon}$,\cite{Weg89}
which in turn implies  $\Delta_q\simeq -\epsilon q(q-1)$. Although the parabolic approximation has turned out to be exact for some models,\cite{LudFSG94} its validity, in particular  for the integer quantum Hall transition, is currently under an intense debate \cite{EveMM08a,ObuSFGL08} due to the implications that this result has upon the critical theories describing the transition.

It is only in the thermodynamic limit where a true critical point exists and hence the true critical $f(\alpha)$ can be obtained. Since the numerical characterization of the multifractal properties of $|\psi|^2$ at the MIT can only be obtained from finite-size states,  one therefore has to consider averages over different realizations of the disorder.  Due to the nature of the distribution of $P_q$,\cite{MirE00,EveM00} one would normally take the {\em typical} average which is exactly the geometric average of the moments of $P_q$  over all contributions.  The use of typical averaging for the MFA has been successfully implemented in various studies.\cite{SchG91,ChaJ89,MilRS97,GruS95}

Remarkably, it was recently argued that an exact {\em symmetry relation} should hold for the anomalous 
scaling exponents,\cite{MirFME06}
\begin{equation}
 	\Delta_q = \Delta_{1-q},
	\label{eq-symDelta}
\end{equation}
which for the singularity spectrum is translated into,
\begin{equation}
 	f(2d-\alpha) = f(\alpha) +d -\alpha.
	\label{eq-symfalfa}
\end{equation}
This relation implies that the singularity strength $\alpha$ must be contained in the interval $[0,2d]$ and that the values of $f(\alpha)$ for $\alpha<d$ can be 
mapped to the values for $\alpha>d$, and vice versa.  We note that the parabolic $f(\alpha)$ \cite{Weg89} is in perfect agreement with this form of the singularity spectrum provided that $f(\alpha)$ is indeed terminated at $0$ and $2d$.
Numerical calculations have since then supported this symmetry in  $f(\alpha)$ in the one-dimensional power-law random-banded-matrix model \cite{MirFME06} and the two-dimensional Anderson transition in the spin-orbit symmetry class.\cite{ObuSFGL07, MilE07}  In the present work we numerically verify that this symmetry in the singularity spectrum also holds in the three-dimensional (3D) Anderson model. In order to address this hypothesis with sufficient accuracy, we have considered the box- and system-size scaling of the typical average of $P_q$ in computing the $f(\alpha)$. We discuss which numerical strategy will produce the best possible agreement with the symmetry and we highlight the statistical analysis that must be used to observe the reported symmetries with sufficient confidence. In a related publication, we also address this problem using the {\em ensemble}-averaged approach\cite{RodVR08} and the reader may wish to compare both articles for a more complete picture of the MFA at the Anderson transition.

\section{The model and its numerical diagonalization}
\label{sec-model}
We use the tight-binding Anderson Hamiltonian in lattice site basis as given by
\begin{equation} \label{anderson_H1} 
\mathcal{H}=\sum_{i} \varepsilon_i~\vert i\rangle\langle i\vert + \sum_{i\neq j} t_{ij}~\vert i\rangle\langle j\vert,
\end{equation}
where site $i=(x,y,z)$ is the position of an electron in a cubic lattice of volume $V=L^3$, $t_{ij}$ are nearest-neighbour hopping amplitudes and 
$\varepsilon_i$ is the random site potential energy.
We consider $\varepsilon_i$ to have a box probability distribution in the interval $[-W_c/2,~W_c/2]$, where $W_c$ is taken to be the strength of the critical value of the disorder.
We assume $W_c=16.5$, above which all eigenstates are localised.\cite{SleMO03,SleMO01,OhtSK99,MilRSU00}
Furthermore the hopping amplitude is taken to be $t=1$  and periodic boundary conditions are used to minimize boundary effects.

\begin{figure}
  \centering
   \includegraphics[width=\figwidth]{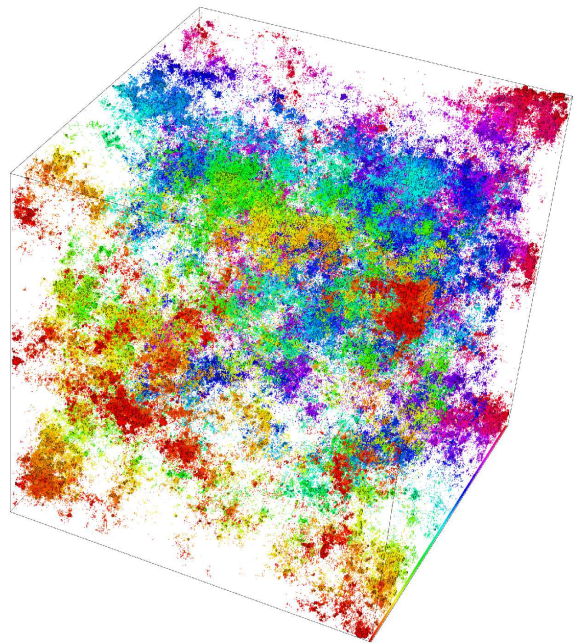}
   \caption{Multifractal eigenstate for the 3D Anderson model at $E=0$  and $W_c = 16.5$ for linear system size $L$=240 with periodic boundary conditions. The $410075$ sites with probability $|\psi_j|^2$ twice larger than the average $1/L^3$ are shown as  boxes with volume $|\psi_j|^2 L^3$. The $26097$ boxes with $|\psi_j|^2 L^3 > 2\sqrt{1000}$ are plotted with black edges. The color/grey scale distinguishes between different slices of the system along the axis into the page.
}
\label{fig-3Dwavefunction}
\end{figure}

The $L^3\times L^3$ Hamiltonian is diagonalized using the JADAMILU package\cite{BolN07} which is a Jacobi-Davidson implementation with an integrated solver based on the incomplete-$LU$-factorization package ILUPACK.\cite{SchBR06, BolN07}
We have considered eigenstates $\psi=\sum_{i} \psi_i |i\rangle$ only in the vicinity of the band centre $E=0$, taking about five eigenstates in a small energy window at $E=0$ for any given realization of disorder.  A list of the number of states and the size of $\psi_i$ used for each $L$ is given in Table \ref{tab-Llist}.  For computing the singularity spectrum using the so-called box-size scaling approach (see section \ref{sec-box}), the largest system size we used is $L=240$ with $95$ eigenstates. This translates into $1.3\times 10^9$ values of wave function amplitudes $\psi_i$. For the system-size scaling, we used all system sizes in Table \ref{tab-Llist}.
A critical eigenstate for $L=240$ is shown in Fig.\ \ref{fig-3Dwavefunction}.

\begin{table}
\caption{The linear system size $L$, corresponding volume $V$ and the number of samples taken and correspondingly the total number of wavefunction amplitudes $\psi_i$ evaluated.  All eigenstates are at the critical MIT for which $W_c=16.5$ and correspond to the five eigenvalues closest to the band center.  The diagonalization of the matrix was computed on an SGI Altix 3700BX2 where for $L=240$ it took approximately 24 hours and requires $\sim$24GB of memory to obtain five eigenstates for one disorder realization.}
\begin{tabular}{r r r r}
\hline
\hline
$L$ & $V=L^3$         & samples  & $\psi_i$ \mbox{    }\\\hline
20  & $8\times 10^3$   &  24 995 &  $2\times 10^8$\\
30  & $9\times 10^3$   &  25 025 &  $6.8\times 10^8$\\
40  & $6.4\times 10^4$ &  25 025 &  $1.6\times 10^9$\\
50  & $1.3\times 10^5$ &  25 030 &  $3.1\times 10^9$\\
60  & $2.2\times 10^5$ &  25 030 &  $5.4\times 10^9$\\
70  & $3.4\times 10^5$ &  24 950 &  $8.6\times 10^9$\\
80  & $5.1\times 10^5$ &  25 003 &  $1.3\times 10^{10}$\\
90  & $7.3\times 10^5$ &  25 005 &  $1.8\times 10^{10}$\\
100 & $1\times 10^6$   &  25 030 &  $2.5\times 10^{10}$\\
140 & $2.7\times 10^6$ &  105    &  $2.9\times 10^8$\\
160 & $4.1\times 10^6$ &  125    &  $5.1\times 10^8$\\
180 & $5.8\times 10^6$ &  100    &  $5.8\times 10^8$\\
200 & $8\times 10^6$   &  100    &  $8\times 10^8$\\
210 & $9.3\times 10^6$ &  105    &  $9.7\times 10^8$\\
240 & $1.4\times 10^7$ &  95     &  $1.3\times 10^9$\\
\hline
\hline
\end{tabular}  
\label{tab-Llist}
\end{table}
%
\section{Multifractal Analysis}
\label{sec-mfa}
%
\subsection{Basic definitions}
\label{ssec-mfa-basic}
Let  $|\psi_i|^2$ be the value at the $i$-th site of a normalized electronic wavefunction in a discretized $d$-dimensional system with volume $L^d$.
If we cover the system with $N_l$ boxes of linear size $l$, the probability to find the electron in the $k$-th box is simply given by 
\begin{equation}
	\mu_k(l)=\sum_{i=1}^{l^d} \vert\psi_i\vert^2,\quad k=1,\ldots,N_l.
	\label{eq-mudef}
\end{equation}
The $\mu_k(l)$ constitutes a normalized measure for which we can define the $q$-th moment as
\begin{equation}
 	P_q (l)=\sum_{k=1}^{N_l}\mu_k^q(l).
	\label{eq-IPRdef}
\end{equation}
The moments $P_q$ can be considered as the generalized inverse-participation ratios (gIPR) for the integrated measure $\mu_k(l)$, reducing to the wave function itself in the case $l=1$ (in units of the lattice spacing) and to the usual IPR $P_2$ for $q=2$. The general assumption underlying multifractality is that within a certain range of values for the ratio $\lambda\equiv l/L$, the moments
$P_q$ show a power-law behaviour indicating the absence of length scales in the system,\cite{Jan94a}
\begin{equation}
	P_q (\lambda)\propto\lambda^{\tau(q)}.
	\label{eq-IPRscale}
\end{equation}
The mass exponent $\tau(q)$ is  defined as
\begin{equation}
	\tau(q) = \begin{cases}
	d (q-1) & \textrm{for metals,}\\
	0 & \textrm{for insulators,}\\
	D_q (q-1) & \textrm{at the MIT.}
	\end{cases}
\end{equation}
The $q$-dependence of the so-called generalized fractal dimensions $D_q$, and therefore a non-linear behaviour of $\tau(q)$,
 is an indication of multifractality. $D_q$ is a monotonically decreasing positive function of $q$ and $D_0$ is equal to the dimension of the support of the measure.
At criticality, $\tau(q)$ can also be parametrized as $\tau(q)=d(q-1)+\Delta_q$, 
where $\Delta_q$ are the anomalous scaling exponents characterizing the critical point. \cite{EveMM08}
The singularity spectrum $f(\alpha)$ is obtained
from the $\tau(q)$ exponents via a Legendre transformation,
\begin{subequations}
\begin{align}
	\alpha_q  &= \frac{\textrm{d}\tau(q)}{\textrm{d}q},\quad q=f^\prime (\alpha), \label{eq-legendre-alfa}\\
	f_q\equiv f(\alpha_q) &= \alpha_q  q-\tau(q). 
\end{align}
\label{eq-legendre}
\end{subequations} 
Here, $f(\alpha)$ denotes the fractal dimension of the set of points where the wavefunction intensity is
$|\psi_i|^2\sim L^{-\alpha}$, that is in our discrete system the number of such points $N_\alpha$ scales as $L^{f(\alpha)}$. \cite{negative}

The singularity spectrum $f(\alpha)$ is a convex function of $\alpha$ and  it has its maximum at $\alpha_0 \geqslant d$ where $f(\alpha_0)=d$.
For $\alpha_1$ we have $f(\alpha_1)=\alpha_1$ and $f^\prime (\alpha_1)=1$. In the limit of vanishing disorder the singularity spectrum becomes narrower and eventually converges to one point $f(d)=d$. On the other hand, as the value of disorder increases the singularity spectrum broadens and 
in the limit of strong localisation the singularity spectrum tends to converge to the points:  $f(0)=0$ and $f(\infty)=d$.
Only at the MIT we can have a true multifractal behaviour and as a consequence the singularity spectrum must be independent of all length scales, such as
the system size.

The symmetry law \eqref{eq-symDelta} can also be written as
\begin{equation}
 	\alpha_q + \alpha_{1-q} = 2d.
\label{eq-symalfaq}
\end{equation}
Since, due to the wave function normalization condition\cite{norm}, the singularity strength $\alpha$ is always positive, it readily follows that the symmetry requires $\alpha\leqslant 2d$. 
Moreover the $0\leqslant \alpha \leqslant d$ and $d\leqslant \alpha \leqslant 2d$ regions of the singularity spectrum must be related by $f(2d-\alpha)=f(\alpha)+d-\alpha$, 
as can be checked by combining Eqs.~\eqref{eq-legendre} and \eqref{eq-symalfaq}.

\subsection{Numerical determination of $f(\alpha)$ at the MIT using typical average}
\label{ssec-computefalpha}
The numerical analysis is essentially based on an \textsl{averaged} form of the scaling law \eqref{eq-IPRscale} in the limit $\lambda\equiv{l}/{L}\rightarrow 0$. This can be achieved either by making the box size $l \rightarrow 0$ for a fixed system size $L$, or by considering $L\rightarrow\infty$ for a fixed box-size. The question of how to compute a proper average of the moments $P_q$ is determined by the form of their distribution function. \cite{MirE00, EveM00}
The scaling law for the typical average of the moments $P_q$ is defined as
\begin{equation}
 	e^{\left<\ln P_q(\lambda)\right>}\propto\lambda^{\tau^{\textrm{typ}}(q)},
 	\label{eq-TYPscale}
\end{equation}
where $\langle\cdots\rangle$ denotes the arithmetic average over all realizations of disorder, i.e.\ over all different wavefunctions at criticality. 
The scaling exponents are then defined by
\begin{equation}
 	\tau^\textrm{typ}(q) = \lim_{\lambda\rightarrow 0}\frac{\langle \ln P_q (\lambda)\rangle}{\ln \lambda},
	\label{eq-tautyp}
\end{equation}
and can be obtained from the slope of the linear fit of $\langle \ln P_q \rangle$ versus 
$\ln\lambda$. Applying Eqs.~\eqref{eq-legendre} we obtain similar definitions for $\alpha$ and $f(\alpha)$,
\begin{subequations}
\begin{align}
	\alpha^\textrm{typ}_q &=  \lim_{\lambda\rightarrow0}\;\frac{1}{\ln\lambda}\left\langle \sum_{k=1}^{N_\lambda}\delta_k(q,\lambda)\ln\delta_k(1, \lambda)\right\rangle 
	\label{eq-alfatyp}, \\
	f^\textrm{typ}_q  &= \lim_{\lambda\rightarrow0}\;\frac{1}{\ln\lambda}\left< \sum_{k=1}^{N_\lambda}\delta_k(q,\lambda)\ln\delta_k(q,\lambda)\right>,
	\label{eq-ftyp}
\end{align}
\label{eq-falfatyp}
\end{subequations}
where $\delta_k(q,\lambda)\equiv \mu_k^q(\lambda) / P_q(\lambda)$ is the normalized $q$-th power of the integrated probability distribution $\mu_k(\lambda)$.
The singularity spectrum could also be obtained from $\tau(q)$ by means of the numerical Legendre transformation \eqref{eq-legendre}, but this latter method 
 is numerically less stable. 

The typical average is dominated by the behaviour of a single (representative) wavefunction. It is because of this
that the $f^\textrm{typ}(\alpha)$ will usually only have positive values, since the average number of points in a single wavefunction with a singularity $\bar{\alpha}$ such that
$f(\bar{\alpha})<0$ is $ L^{-|f(\bar{\alpha})|} \ll 1$. It is also worth mentioning that due to the relation \eqref{eq-legendre-alfa}, the typical singularity spectrum is expected to approach the abscissa axis with an infinite slope. However, it has been proven numerically, that the region of $\alpha$ values near the ends where the slope tends to diverge gets narrower and eventually disappears as the thermodynamic limit is approached. \cite{EveMM01} 

\section{Scaling with box size}
\label{sec-box}

In the scaling law of Eq.~\eqref{eq-IPRscale}, the limit $\lambda\rightarrow 0$ can be reached by taking the box size $l\rightarrow 0$, i.e.\ we are  evaluating the scaling of $P_q$ with box size $l$ at constant $L$. Numerically, we consider a system with large $L$ and we partition it into smaller boxes such that condition $ l_m\ll l<L$ is fullfilled with $l_m$ the lattice spacing. This ensures that the multifractal fluctuations of $|\psi|^2$ will be properly measured. We usually take  values of the box size in the range  $l\in[10,L/2]$.
We have found that the most adequate box-partitioning scheme is when the system is divided into \textit{integer} number of \textit{cubic} boxes, each box with linear size $l$.\cite{VasRR08}
The system is partitioned in such a way that it can be divided equally into boxes and the origin of the first box coincides with the origin $(x,y,z)=(0,0,0)$ of the system.
We have used this method to produce all the results in this section.
We have also tried other box-partitioning strategies, however, their results were less accurate and will be discussed elsewhere.\cite{VasRR08} 

\begin{figure}
  \centering
  \includegraphics[width=\figwidth]{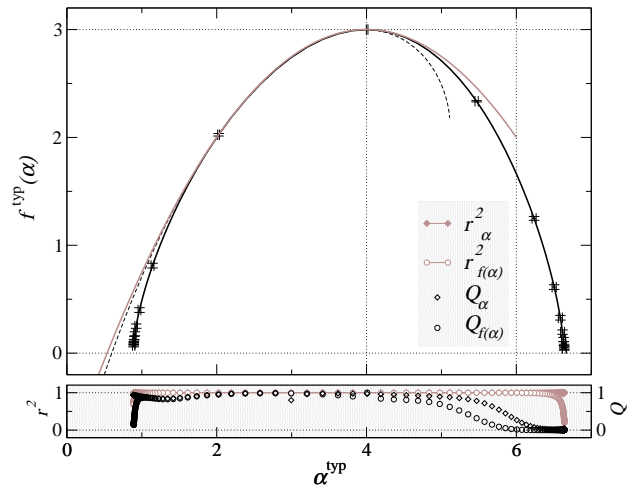} 
  \caption{
Singularity spectrum (thick solid black line) obtained using box-size scaling of the typical average of $P_q$ for  
system size $L = 240$ with $95$ states.  The error bars which are equal to one standard deviation mark the locations corresponding to integer $q$ values.
The corresponding symmetry-transformed spectrum \eqref{eq-symfalfa} is shown as thin black dashed line.  The analytical parabolic form \cite{Weg89} is represented by the thick gray solid line.  The thin horizontal and vertical lines denote the $f=0$, $f=3$ and $\alpha=4$, $\alpha=6$ values, respectively.  The values for the linear correlation coefficient $r^2$ and quality-of-fit parameter $Q$ for both $\alpha^\textrm{typ}$ and $f^\textrm{typ}(\alpha)$  are shown in the bottom shaded panel.}
\label{fig-3Dbox-typ-best}
\end{figure}

\begin{figure}
  \centering
  \includegraphics[width=\figwidth]{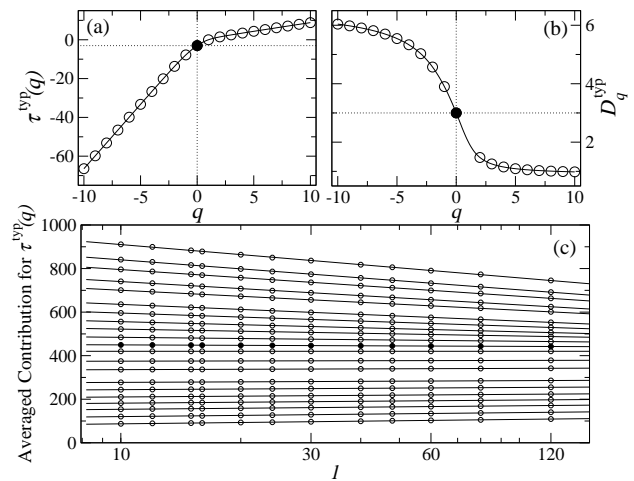} 
  \caption{Mass exponents (a) and generalized fractal dimensions (b) obtained using box-size scaling  of the typical average of $P_q$ for $L=240$ considering $95$ states. The filled black circles correspond to $q=0$ while the empty circles denote integer $q$.  The thin horizontal and vertical lines indicate the $\tau(0)=-3$, $D_0=3$ and $q=0$ values, respectively.  In panel (c), we show the linear fits of Eq.~\eqref{eq-tautyp} for the $\tau^\textrm{typ}(q)$ in panel (a). Only fits for integer values of $q$ ranging from $q=-10$ (top) to $q=10$ (bottom) are shown. The values of $\tau^\textrm{typ}(q)$ are given by the slopes of the fits. Data points for $q\neq 0$ have been properly shifted vertically to ensure optimal visualization.  Data for $q=0$  is highlighted with filled symbols. Standard deviations of all data are contained within symbol size in all cases.}
\label{fig-3Dbox-typ-TauqDq}
\end{figure}

For each wave function, we compute for the $q$-th moment of the box probability in each box, and $P_q$ as in \eqref{eq-IPRdef}, as its sum from all the boxes.  The scaling behaviour of the averaged gIPR  with box size \eqref{eq-TYPscale} is then obtained by varying $l$.  Finally, the corresponding values of the singularity strength $\alpha$ and spectrum $f(\alpha)$ are derived from the linear fits of the Eqs.~\eqref{eq-falfatyp} in terms of the box size.  With only one system size to be considered, the box-size scaling is numerically relatively inexpensive and has been much used previously in performing a MFA.\cite{SchG91,MilRS97,GruS95} In Figs.\ \ref{fig-3Dbox-typ-best} and \ref{fig-3Dbox-typ-TauqDq} we show examples of $f(\alpha)$ and associated linear fits.
%
\subsection{General features of $f^\textrm{typ}(\alpha)$}

The singularity spectrum for system size $L=240$ with $95$ states that is obtained using Eqs.~\eqref{eq-falfatyp} with $l\rightarrow 0$ is shown in Fig.~\ref{fig-3Dbox-typ-best}.  The $f^\textrm{typ}$ is compared with the corresponding spectrum that is derived from the symmetry relation \eqref{eq-symfalfa} and with the parabolic spectrum.\cite{Weg89}  
Here, the maximum $f^{\textrm{typ}}(\alpha_0)=3$ which is equal to the dimension of the support can be found very near to $\alpha_0=4$ where the maximum  of the parabolic spectrum is located at.\cite{Weg89}  In the region within the vicinity of $\alpha=3$, the typical singularity spectrum closely resembles the parabolic $f(\alpha)$.  However, for large $\vert q\vert$ values particularly at the tails, the $f^{\textrm{typ}}(\alpha)$ starts to deviate from the parabolic spectrum.  We note that the symmetry relation \eqref{eq-symfalfa} requires that the spectrum should be contained below the upper bound of $\alpha=2d$.

In order to obtain $\alpha$ and $f(\alpha)$ via the linear fit of Eqs.~\eqref{eq-falfatyp}, a general $\chi^2$ minimization is considered taking into account the statistical uncertainty of the averaged right-hand side terms. In this way we can carry out a complete analysis of the goodness of the fits via the quality-of-fit parameter $Q$, as well as the usual linear correlation coefficient $r^2$.  The behaviour of these quantities for the different parts of the spectrum (corresponding to different values of the moments $q$) is shown in the bottom panel of Fig.~\ref{fig-3Dbox-typ-best}.  The $r^2$ value is almost equal to one for all $\alpha$ which shows the near perfect linear behaviour of the data.  Furthermore,  acceptable values of the $Q$ parameter are also obtained. However, a decline in the $r^2$ and $Q$ values is seen at the tails. These regions correspond to the large $\vert q\vert$ values where the numerical uncertainties in computing for the $P_q$ over a number of different disorder realizations are large enough to affect the reliability of the data. 
Figure \ref{fig-3Dbox-typ-TauqDq} presents the corresponding sets of mass exponents $\tau^\textrm{typ}(q)$, generalized fractal dimensions $D^\textrm{typ}_q$ and linear fits for $\tau^\textrm{typ}(q)$ for the singularity spectrum in Fig.~\ref{fig-3Dbox-typ-best}. 
The $q$-dependence of the decreasing function $D_q\equiv \tau(q)/(q-1)$ is an indication of multifractality.  Here, we see that $D_0=d$ as expected.  The corresponding $\tau^\textrm{typ}(q)$  is shown in Fig.~\ref{fig-3Dbox-typ-TauqDq}(a).  It displays the characteristic nonlinearity of a multifractal where $\tau(0)=-d$.  The regions corresponding to large $\vert q\vert$ values show a linear behaviour with a constant slope.  Since the singularity strength is defined as $\alpha_q = \tau'(q)$ then a linearity in $\tau^\textrm{typ}(q)$ found in the limit of $\vert q\vert\rightarrow\infty$ results in $\alpha^\textrm{typ}_q$ values that approach upper $\alpha_+$ and lower $\alpha_-$ bounds.
Hence, the $f^\textrm{typ}(\alpha)$ meets the $\alpha^\textrm{typ}$ axis at these termination points with an infinite slope.  Furthermore, we will show that the location of $\alpha_-$ and $\alpha_+$ is greatly affected by system size.
For a detailed discussion on the relationship between the shapes of $\tau(q)$ and $f(\alpha)$, we refer to the references \onlinecite{EveMM08} and \onlinecite{EveMM01}.

\begin{figure}
  \centering
  \includegraphics[width=\figwidth]{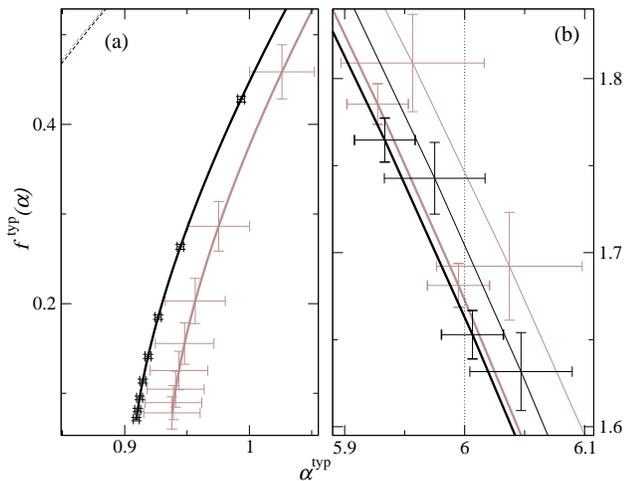}
  \caption{Singularity spectrum obtained using box-size scaling of the typical average of $P_q$.  
In panel (a): system size $L=90$ for $250$  (thick gray line)  and $2.5\times 10^4$ (thick black lines) states. The corresponding symmetry-transformed spectra \eqref{eq-symfalfa} are shown as thin dashed lines. The error bars mark the location of integer $q$ values. In panel (b): system size $L=60$ (thick gray line), $L=100$ (thin gray line), $L=200$ (thin black line) and $L=240$ (thick black line) each having $10^3$, $10^2$, $10^2$ and $95$ states respectively. The dashed vertical line indicates $\alpha=6$. In all cases, the error bars denote one standard deviation.}
\label{fig-3Dbox-typ}
\end{figure}

\subsection{Effects of the number of states and $L$}

In panel (a) of Fig.~\ref{fig-3Dbox-typ}, we show the small-$\alpha$ region of $f^\textrm{typ}(\alpha)$ for the case of $L=90$ with $2.5\times10^2$ and $2.5\times10^4$ states.  When more states are considered for a fixed system size, the termination point $\alpha_-$ moves further towards smaller values (i.e., towards the thermodynamic limit) and the symmetry relation is more closely satisfied. 
However, when a large number of states has already been considered (such as $2.5\times10^4$ for $L=90$) the shape of the $f^\textrm{typ}(\alpha)$  will not significantly change anymore with more states as illustrated by the already small uncertainties. 
This takes us to consider bigger system sizes in order to be able to improve the symmetry relation.
In panel (b) of Fig.~\ref{fig-3Dbox-typ}, we show a portion of the large-$\alpha$ part of $f^\textrm{typ}(\alpha)$ for varying system sizes $L=60$ with $10^3$ states, $L=100$ and $L=200$ with $10^2$ states each, and $L=240$ with $95$ states.
We see that for the same number of states the degree of fluctuations as represented by the size of the error bars is larger for smaller system size.  
Moreover, the $f^\textrm{typ}(\alpha)$ for $L=60$ with $10^3$ states is, within the standard deviations, the same as that for $L=240$ with $95$ states.  
This can be explained by the total number of wavefunction values $\psi_i$ involved in the average, which are nearly the same for both cases and hence causes the same shape of $f^\textrm{typ}(\alpha)$. This means that when using box-size scaling for the typical average of $P_q$, the number of disorder realizations needed to obtain the singularity spectrum up to a given degree of reliability decreases with the size of the system.  
Remarkably, we also see in Fig.~\ref{fig-3Dbox-typ}(b) a general tendency that with larger $L$ the singularity spectrum approaches the upper bound of $\alpha=6$ in keeping with what the symmetry relation requires.

\begin{figure}
  \centering
  \includegraphics[width=\figwidth]{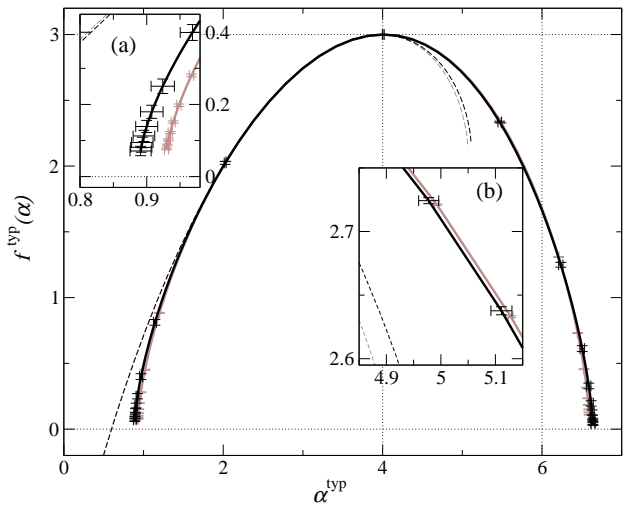}  
  \caption{Singularity spectrum obtained using box-size scaling of the typical average of $P_q$, for 
system sizes $L=60$ (thick gray line) and $L=240$ (thick black line) with $2.5\times 10^4$ and $95$ states respectively.  
The corresponding symmetry-transformed spectra \eqref{eq-symfalfa} are shown as thin dashed lines. The insets show details for (a) small and (b) large $\alpha$ values. In all cases, the error bars are equal to one standard deviation.  In inset (a), the error bars mark the location corresponding to integer $q$ values.}
\label{fig-3Dbox-typ-L}
\end{figure}
In Fig.~\ref{fig-3Dbox-typ-L}, we show the spectra corresponding to  $L=60$ with $2.5\times 10^4$ and $L=240$ with $95$ states to clearly show the effect of the system size. We observe that the value of $\alpha_0$ (i.e., location of the maximum) and the shape of the singularity spectrum near the maximum do not change anymore with $L$.  This $L$-invariant behaviour of the singularity spectrum is an attribute of a critical point. 
In inset (a), with increasing system size the position of the termination point $\alpha_-$ moves towards smaller values.  Furthermore, a closer look of $f^\textrm{typ}(\alpha)$ in insets (a) and (b) shows that when a bigger system size is used, even with less eigenstates,  there is a well defined improvement to satisfying the symmetry.

\subsection{Symmetry relation}

\begin{figure}
  \centering
  \includegraphics[width=\figwidth]{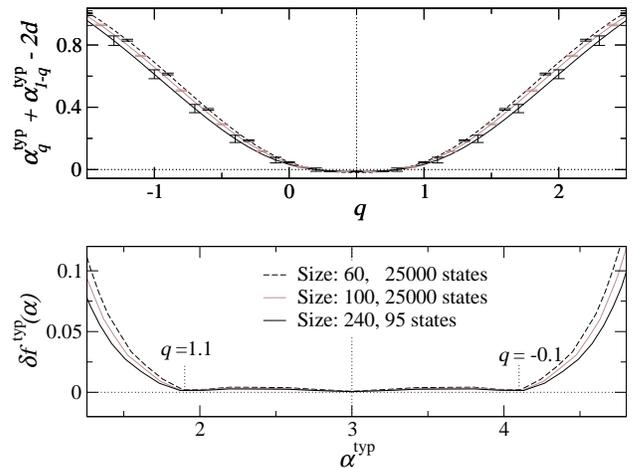}
  \caption{Measures for the degree of symmetry of the multifractal spectrum obtained from the typical average with box-size scaling. The upper panel shows the  numerical evaluation of the symmetry law as a function of $q$ for system sizes $L=60$ with $2.5\times 10^4$ states (dashed black), $L=100$ with $2.5\times 10^4$ (dark grey) and $L=240$ (black) with $95$ states.
The bottom panel shows $\delta f(\alpha)$ versus $\alpha$. Note that there is no correspondence between the abscissa axes of the upper and lower plots. For clarity, two values of $q$ for the black line are explicitly written.} 
\label{fig-3Dbox-typ-symdeg}
\end{figure}

In order to quantify how the symmetry is being satisfied with regards to either taking more states or considering bigger system size, we present 
Fig.~\ref{fig-3Dbox-typ-symdeg}.  The top panel is an exact calculation of the symmetry relation of Eq.~\eqref{eq-symalfaq} whereas the bottom panel shows the difference between the singularity spectrum and its symmetry-transformed counterpart, defined as 
\begin{equation}
 	\delta f(\alpha) \equiv | f(\alpha)-f(2d-\alpha)+d-\alpha|.
	\label{eq-fdistance}
\end{equation}
The latter plot is an effective tool to tell us the range of the $\alpha$ values where the symmetry is satisfied up to a given tolerance. However, directly comparing the degree of symmetry via $\delta f(\alpha)$ is just an approximation since (i) linear interpolation has to be used to measure the vertical distance properly at several values of $\alpha$, and (ii) for a given $q$ the corresponding value of $\alpha$ as well as its uncertainty depend upon sizes and realizations of disorder, and this makes the comparison of the different curves in terms of $\alpha$ not as reliable as Eq.~\eqref{eq-symalfaq}. In fact, the resulting error bars are much larger than in the top panel of Fig.~\ref{fig-3Dbox-typ-symdeg} and even larger than the variation between the 3 shown $\delta f(\alpha)$ curves. Nevertheless, the results in Fig.~\ref{fig-3Dbox-typ-symdeg} illustrate that there is a tendency to find a better agreement with the symmetry relation whenever more states or bigger system sizes are considered. The best situation corresponds to the biggest system size available ($L=240$) even though the number of eigenstates is lower than for smaller systems.  
The relatively weak effect of the number of states on the shape of the singularity spectrum is a result of taking the typical average where by nature the average does not dramatically change with the number of samples taken. Furthermore, a rough estimation from our results suggest that in order to obtain numerically a good $f(\alpha)$ symmetry relation ($\delta f(\alpha) \leqslant 0.01$) for $\alpha \in [1.5, 4.5]$ using box-size scaling one would have to consider very big system sizes $L\gg 1000$. 

\section{Scaling with system size}
\label{sec-size}

The scaling law of the gIPR  \eqref{eq-IPRscale} can also be studied in terms of the system size $L$.
Obviously the numerical calculation of eigenstates for very large 3D systems is a demanding task. \cite{ElsMMR99,ZhaK99,SchMRE99,SchBR06}
Hence previous  MFA studies at the MIT  have been mostly based on the box-partitioning scaling described in Sec.~\ref{sec-box}.
One naturally would expect the scaling with the system size to perform better 
in revealing the properties of the system  in the thermodynamic limit. 
The fact that for each system size one has several independent realizations of the disorder helps reduce finite-size effects, which 
will be unavoidably more pronounced when doing scaling with the box size. 
Obviously the larger the system sizes and the more realizations of the disorder, the better.

\subsection{Coarse-graining for negative $q$}
\label{sec-coaese}

In the present case the scaling variable is $L$, and the formulae \eqref{eq-tautyp} and \eqref{eq-falfatyp} for the singularity spectrum are only affected by the substitution: 
$\lim_{\lambda\rightarrow 0}\Rightarrow -\lim_ {L\rightarrow\infty}$. The box size $l$ which determines the integrated probability distribution $\mu_k(l)$ is  now a parameter in the expressions  \eqref{eq-tautyp} and \eqref{eq-falfatyp} for $\tau^\textrm{typ}(q),\, \alpha^\textrm{typ}_q$ and $f^\textrm{typ}_q$. Changing the value of $l$ is effectively equivalent to renormalize the system size to a smaller 
value $L^\prime \equiv L/l$. Therefore it is clear that the most favourable situation to approach the thermodynamic limit is setting $l=1$, thus defining the generalized 
IPR in terms of the wavefunction itself, $P_q=\sum_{i=1}^{L^3}|\psi_i|^{2q}$. However, when considering negative moments, all the possible numerical inaccuracies that 
may exist in the small values of $|\psi_i|^2$ will be greatly enhanced, which in turn causes a loss of precision in the right branch ($\alpha>\alpha_0$) of the singularity spectrum. The best way to fix this problem is to use a box-size $l>1$ for $q<0$. In this way the relative uncertainties in the smallest values of the coarse-grained integrated distribution $\mu_k(l)$ are reduced with respect to the values of the wavefunction. This coarse-graining procedure to evaluate the negative moments of the wavefunction when doing system-size scaling was first described  in Ref.~\onlinecite{MirFME06} and as we have seen its validity is readily proven when one assumes the scaling relation \eqref{eq-IPRscale} as the starting point of the MFA. 

The numerical singularity spectrum is thus obtained from the slopes of the linear fits in the plots of the averaged terms in Eqs.~\eqref{eq-falfatyp} 
versus $\ln L$, for different values of the system size $L$. Where for positive $q$ we 
have $\mu_k(1)=|\psi_k|^2$ and for negative $q$ the integrated measure $\mu_k(l>1)$ is kept, with $l=5$ in most of the calculations. 
The value of $l$ for the coarse-graining procedure should not be very large, otherwise finite-size effects will be enhanced again due to the reduction in the effective system size. 
For the benefit of the reader let us rewrite the formulae \eqref{eq-falfatyp} in the particular case where $l=1$,
\begin{subequations}
\begin{align}
	-\alpha^\text{typ}_q  \ln L & \sim  \left\langle \frac{\sum_i |\psi_i|^{2q} \ln |\psi_i|^2}{\sum_j |\psi_j|^{2q}} \right\rangle , \\
	- f^\text{typ}_q \ln L & \sim  \left\langle \frac{\sum_i |\psi_i|^{2q} \ln |\psi_i|^{2q}}{\sum_j |\psi_j|^{2q}} 
	-  \ln \sum_i |\psi_i|^{2q}\right\rangle ,
\end{align}
\label{eq-falfa-typ-psi}
\end{subequations}
for large enough system sizes $L$. As before the angular brackets denote the average over all eigenstates. 

\subsection{General features of $f^\textrm{typ}(\alpha)$ and the effects of the number of states and $L$}
\begin{figure}
  \centering
  \includegraphics[width=\figwidth]{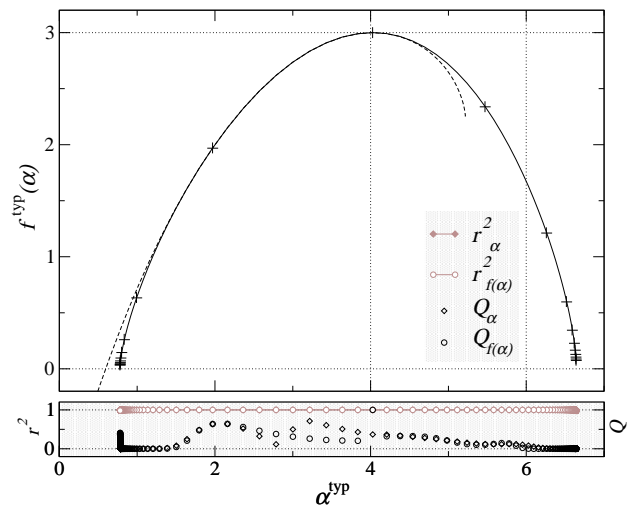}
  \caption{Singularity spectrum obtained from typical averaging. System sizes from $L=20$ to $100$ have been considered with $\sim2.5\times10^4$ different wavefunctions 
	for each system size, as shown in Table \ref{tab-Llist}. The dashed line is the symmetry-transformed spectrum according to 
	$f(6-\alpha)=f(\alpha) +3 -\alpha$. The values of $q$ range from $q=-10$ to $q=10$ with a step of $0.1$ ($l=1$ for $q \geqslant 0$ and $l=5$ for $q<0$). Error bars which are equal to one standard deviation highlight the values corresponding to integer $q$. The lower panel shows the linear correlation coefficient ($r^2$) and the quality-of-fit parameter ($Q$) of the linear fits to obtain the values for $\alpha$ and $f(\alpha)$.}
\label{fig-3Dsize-typ}
\end{figure}

\begin{figure}
  \centering
  \includegraphics[width=\figwidth]{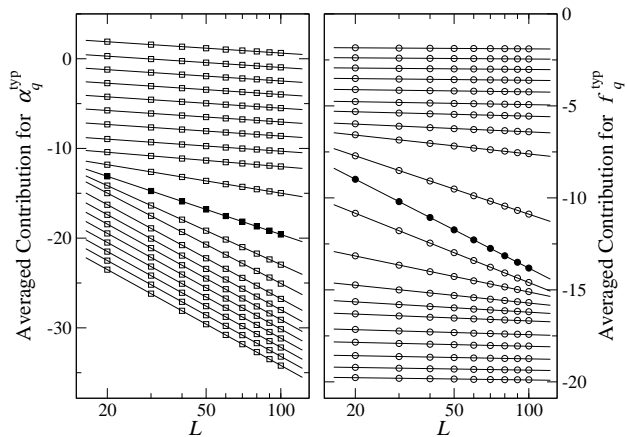}
  \caption{Linear fits of Eqs.~\eqref{eq-falfatyp} for  $\alpha^\textrm{typ}_q$ values (left) and $f^\textrm{typ}_q$ values (right) of the singularity spectrum in Fig.~\ref{fig-3Dsize-typ}. Only fits for integers values of $q$ ranging from $q=10$ (top) to $q=-10$ (bottom) are shown. The values of $\alpha^\textrm{typ}_q$ and 
$f^\textrm{typ}_q$ are given by the slopes of the fits. Data points for $q\neq 0$ have been properly shifted vertically to ensure optimal visualization.  Data for $q=0$ highlighted with filled symbols. Standard deviations are contained within symbol size in all cases.}
\label{fig-3Dsize-typ-fits}
\end{figure}

\begin{figure}
  \centering
  \includegraphics[width=\figwidth]{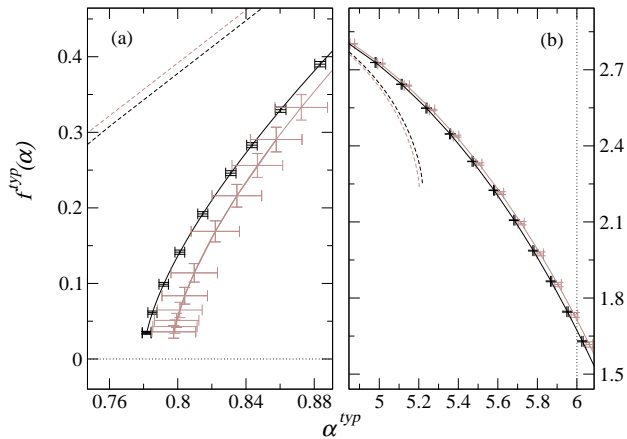}
  \caption{Left branch (a) and right branch (b) of the singularity spectrum obtained from typical average, scaling with system sizes from $L=20$ to $100$ for $\sim2.5\times 10^4$ states (black) and $10^3$ (grey) for each system size. Dashed lines correspond to spectra transformed according to the symmetry law.  The values of $q$ range from $q=-10$ to $q=10$ with a step of $0.1$ ($l=1$ for $q \geqslant 0$ and $l=5$ for $q<0$). Error bars are standard deviations.}
\label{fig-3Dsize-typ-stateseffect}
\end{figure}

In Fig.~\ref{fig-3Dsize-typ} we show the singularity spectrum obtained from Eqs.~\eqref{eq-falfatyp} ($q < 0$) and \eqref{eq-falfa-typ-psi} ($q\geqslant 0$). 
We have considered system sizes ranging from $L=20$ to $100$, and $\sim2.5\times 10^4$ states for each system size, as shown in Table \ref{tab-Llist}. In spite of the good linear behaviour observed in the fits to obtain $\alpha^\textrm{typ}_q$ and $f^\textrm{typ}_q$, shown in Fig.~\ref{fig-3Dsize-typ-fits}, the values for $Q$ in the bottom panel of Fig.~\ref{fig-3Dsize-typ}, suggest a loss of reliability near the termination regions of the spectrum.
On the other hand the standard deviations of the $\{\alpha^\textrm{typ}, f^\textrm{typ}(\alpha)\}$ values are really small even near the ends. These uncertainties are directly related to the number of states we average over: the more realizations, the smaller these uncertainties are. It must be clear that these standard deviations  only give an idea about the reliability of data as a function of the number of disorder realizations for the particular range of system sizes that one is using. To illustrate the influence of the number of disorder realizations upon the typical average a comparison can be found in Fig.~\ref{fig-3Dsize-typ-stateseffect}, between the $f^\textrm{typ}(\alpha)$ spectrum obtained after averaging over $10^3$ states for each system size and the one for $\sim2.5\times 10^4$ states. As can be seen, after this increase in the number of states the overall change in the spectrum is not very significant, altough some variation can be noticed in the regions shown. In particular, the right branch of the spectrum moves inwards and the end of the left tail shifts to smaller values of $\alpha$. In both regions the expected variation of the spectrum is well described by the standard deviations. 
In the case of Fig.~\ref{fig-3Dsize-typ}  according to the standard deviations the conclusion is that a further increase of the number of states will not  mean a  significant change in the shape of the spectrum. Nevertheless it must also be very clear that if we consider a different range of larger system sizes, noticeable changes could happen in the singularity spectrum. The standard deviations do never account for the effects stemming from the range of system sizes used.

\begin{figure}
  \centering
  \includegraphics[width=\figwidth]{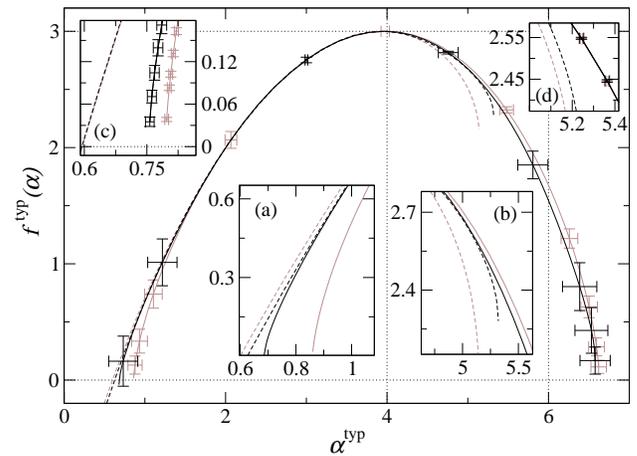}
  \caption{Singularity spectrum obtained from typical average using different ranges of system sizes. Grey line: $7$ system sizes from $L=40$ to $100$ and $10^2$ states for each. Black line: $6$ system sizes from $L=140$ to $240$ and $\sim10^2$ states for each. 
Insets (c) and (d):  Grey line: $5$ system sizes from $L=20$ to $60$ and $\sim2.5\times 10^4$ states for each. Black line: $5$ system sizes from $L=60$ to $100$ and $\sim2.5\times 10^4$ states for each. In all cases dashed lines correspond to spectra transformed via the symmetry law. The values of $q$ range from $q=-10$ to $q=10$ with a step of $0.1$ ($l=1$ for $q \geqslant 0$ and $l=5$ for $q<0$). Error bars are standard deviations.}
\label{fig-3Dsize-typ-lengtheffect}
\end{figure}

To evaluate the effects due to the system size, we compare in Fig.~\ref{fig-3Dsize-typ-lengtheffect} the multifractal spectrum obtained considering different ranges of system sizes with a similar number of disordered realizations. In the main plot it can be seen how the shape of the spectrum changes in its right (large $\alpha$) branch, which moves inwards, when we consider system sizes in the interval $[140,240]$ compared to the situation for sizes in $[40,100]$. The left end of the spectrum also shifts to smaller values of $\alpha$ when larger system sizes are considered. In this case the standard deviations are noticeable since we have only considered $10^2$ states for each system size. In the insets (c) and (d) within Fig.~\ref{fig-3Dsize-typ-lengtheffect} a similar comparison can be found for ranges of smaller sizes, $[60,100]$ versus $[20,60]$ but with a much higher number of states, $\sim2.5\times 10^4$ for each size. In this situation the change is less dramatic, but the tendency remains the same. In particular it should be noticed in Fig.~\ref{fig-3Dsize-typ-lengtheffect}(c) how the change in the left end of the spectrum is not contained in the uncertainty regions given by the error bars, confirming the fact that these standard deviations do not fully describe system size effects. 

\subsection{Symmetry relation}

The symmetry relation \eqref{eq-symfalfa} is only partially fulfilled in Fig.~\ref{fig-3Dsize-typ}. Still, a nice overlap between the original spectrum and the symmetry-transformed one occurs in the region around the symmetry point $\alpha=3$. The agreement is lost when approaching the tails, which are the parts more affected by numerical inaccuracies and system-size effects. For a given range of system sizes, the symmetry relation tends to be better satisfied whenever the number of disordered realizations is increased, as can be observed in Fig.~\ref{fig-3Dsize-typ-stateseffect}. On the other hand the improvement of the symmetry is even more dramatic when we consider larger system sizes to do the scaling, as shown in the insets (a) and (b) of Fig.~\ref{fig-3Dsize-typ-lengtheffect}. In this figure it is evident how the value of $f(\alpha=6)$ decreases when considering larger system sizes, hence tending towards the upper bound at $\alpha=2d$ as predicted by \eqref{eq-symfalfa}.
\begin{figure}
  \centering
  \includegraphics[width=\figwidth]{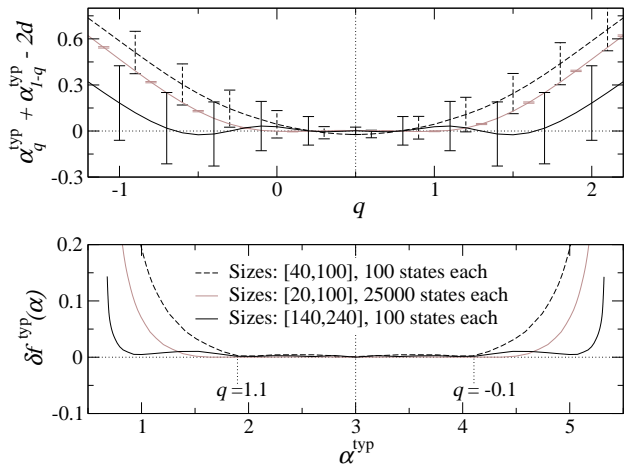}
  \caption{Measure of degree of symmetry of the multifractal spectrum of Fig.\ \ref{fig-3Dsize-typ-lengtheffect} obtained from scaling with system size. The upper panel shows the  numerical evaluation of the symmetry law as a function of $q$. The bottom panel contains $\delta f(\alpha)$ versus $\alpha$. Dashed black: $7$ system sizes from $L=40$ to $100$ and $10^2$ states for each. Grey: $9$ system sizes from $L=20$ to $100$ and $\sim2.5\times 10^4$ states for each. Solid black: $6$ system size from $L=140$ to $240$ and $\sim10^2$ states for each. There is no correspondence between the abscissa axes of the upper and lower plots. For clarity, two values of $q$ for the black line are explicitly written.}
\label{fig-3Dsize-typ-symdeg}
\end{figure}

A quantitative analysis of the symmetry relation is shown in the upper panel of Fig.~\ref{fig-3Dsize-typ-symdeg}. The best data correspond to the scaling with system sizes in $[140,240]$ after averaging over $\sim 100$ states for each size (cp.\ Table \ref{tab-Llist}). Even with such a low number of disorder realizations, the observed symmetry is better on average than the one obtained for sizes in $[20,100]$ with $2.5\times 10^4$ states for each $L$. Let us emphasize that for $L=100$ the total number of wave function values involved in the calculation is $2.5\times 10^{10}$ while for $L=240$ it is only  $1.3\times 10^9$. 
This shows that although the number of disorder realizations is important to improve the reliability of data (reducing the standard deviations), the effect of the range of system sizes is more significant. And although it can be argued that the error bars of the black line in the upper panel of Fig.~\ref{fig-3Dsize-typ-symdeg} are still very large, we have already shown that when increasing the number of states the symmetry simply gets better and thus the line will move even closer to zero. In the lower panel of Fig.~\ref{fig-3Dsize-typ-symdeg} the deviation from symmetry $\delta f(\alpha)$ defined in \eqref{eq-fdistance} is also shown and corroborates these findings.

Hence, whenever the reliability of data is improved by increasing the number of disorder realizations, or when finite-size effects are reduced by considering larger system sizes, we get a better agreement with the symmetry law \eqref{eq-symfalfa} of the multifractal spectrum. Assuming the degree of symmetry is a qualitative measure of the MFA itself, then from a numerical viewpoint, the best strategy when doing scaling with system size and typical averaging would be to go for the largest system sizes accessible even though it means having less realizations of disorder. 
%
\section{Summary and Conclusions}
\label{sec-concl}
We have obtained the multifractal spectrum from the box- and system-size scaling of the typical average of the gIPR. We find that, upon increasing either the number of disorder realizations or by taking larger system size, the 
$f^\textrm{typ}(\alpha)$ spectrum becomes evermore close to obeying the proposed symmetry relation \eqref{eq-symfalfa}. Using the typical average, the best symmetry in the singularity spectrum is obtained by taking large system sizes. Due to the nature of the typical averaging, taking more states only changes the shape of the $f^\textrm{typ}(\alpha)$ up to a point.
By considering larger system sizes, a significant improvement of the symmetry relation is achieved, leading to lower values of $\alpha^\textrm{typ}$ and $f^\textrm{typ}$ on the left side of the spectrum as well as a better agreement with the upper cut-off of $\alpha\leq 6$.

\begin{figure}
  \centering
  \includegraphics[width=\figwidth]{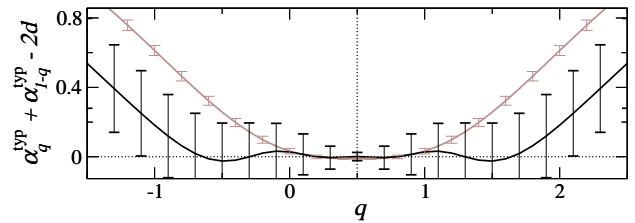}
  \caption{Comparison of degree of symmetry for the spectra obtained using typical averaging of $P_q$ for the cases of box-size [BS] scaling (gray) and system-size [SS] scaling (black).  The best spectrum for each case has been considered: [BS] $L=240$ with 95 states and [SS] $L\in [140,240]$ with 100 states for each size.  The plot shows the  numerical evaluation of the symmetry law as a function of $q$.}
\label{fig-3Dtyp-symdeg}
\end{figure}

In Fig.~\ref{fig-3Dtyp-symdeg}, let us now compare box- and system-size scaling. With system-size scaling the symmetry is (nearly) satisfied for a wider range of $\alpha$ values as compared with the box-size scaling. Box-size scaling is more strongly influenced by finite-size effects. 
However, the agreement with the symmetry relation is lost for both methods at large $\vert q \vert$ or equivalently at $\vert\alpha -3 \vert\gg 0$. Unsurprisingly, these are the regions greatly affected by numerical inaccuracies and finite-size effects. Hence we conclude that within the accuracy of the present calculation and within the limits of the typical averaging procedure, the proposed symmetry relation \eqref{eq-symfalfa} is valid at the Anderson transition in 3D.\cite{PLRBM}

Last, let us remark that the relation \eqref{eq-symfalfa} implies negative values of $f$ for small values of $\alpha$. As discussed previously, this is hard to see using the typical averaging procedure. In Ref.\ \onlinecite{RodVR08}, we have also performed MFA using the {\em ensemble}-averaged box- and system-size scaling approaches. The results again support the existence of the symmetry \eqref{eq-symDelta} for an even larger range of $\alpha$ values and including a negative $f(\alpha)$ part for small $\alpha$.
\acknowledgments
We thank F.\ Evers for a discussion. 
RAR gratefully acknowledges EPSRC (EP/C007042/1) for financial support. AR acknowledges financial support from the Spanish government under contracts 
JC2007-00303, FIS2006-00716 and MMA-A106/2007, and JCyL under contract SA052A07.
\vspace*{10mm}

%
%

\begin{thebibliography}{99}
\bibitem{negative} It is worth mentioning that the values of $f(\alpha)$ are not restricted to be positive,\cite{Man03} in fact a negative fractal dimension has a straightforward interpretation: the negative values of $f(\alpha)$ would correspond to the regions of values of the wavefunction that become more and more empty when approaching the 
thermodynamic limit. That is if $f(\bar{\alpha})<0$ then the number of points where $|\psi_i|^2\sim L^{-\bar{\alpha}}$ will 
decrease with the system size as $L^{-|f(\bar{\alpha})|}$
\bibitem{norm} The most extremely localized state is $\vert \psi_{i}\vert^2 = 1$ at the site $i$ and zero otherwise. Hence $\vert\psi_{i}\vert^2 \propto L^{-\alpha}$ implies $\alpha=0$ as lower boundary.
\bibitem{PLRBM} Using the 1D power-law-random-banded matrix model, we have produced similar results as in the present paper. These results confirm the validity of the relation \eqref{eq-symfalfa} as presented in Ref.\ \onlinecite{MirFME06}.
\end{thebibliography}
%

\end{document}